\documentclass[twocolumn,tighten,times]{aastex62} 
\usepackage{newtxmath} 

\received{June 7, 2019}
\revised{January 20, 2020}
\accepted{February 24, 2020}

\newcommand{\angstrom}{{$\rm \mathring A$}}

\shorttitle{On the puzzling large UV to X-ray lags}
\shortauthors{Cai et al.}

\begin{document}

\title{EUCLIA. II. On the puzzling large UV to X-ray lags in Seyfert galaxies}

\email{zcai@ustc.edu.cn, jxw@ustc.edu.cn}

\author[0000-0002-4223-2198]{Zhen-Yi Cai}
\affiliation{CAS Key Laboratory for Research in Galaxies and Cosmology, Department of Astronomy, University of Science and Technology of China, Hefei 230026, China}
\affiliation{School of Astronomy and Space Science, University of Science and Technology of China, Hefei 230026, China}

\author[0000-0002-4419-6434]{Jun-Xian Wang}
\affiliation{CAS Key Laboratory for Research in Galaxies and Cosmology, Department of Astronomy, University of Science and Technology of China, Hefei 230026, China}
\affiliation{School of Astronomy and Space Science, University of Science and Technology of China, Hefei 230026, China}

\author[0000-0002-0771-2153]{Mouyuan Sun}
\affiliation{Department of Astronomy, Xiamen University, Xiamen, Fujian 361005, China}


\begin{abstract}

Recent intense X-ray and UV monitoring campaigns with {\it Swift} have detected clear UV lags behind X-ray in several local active galactic nuclei (AGNs). The UV to X-ray lags are often larger (by a factor up to $\sim$20) than expected if the UV variation is simply due to the X-ray reprocessing. We previously developed a model in which the UV/optical variations are attributed to disk turbulences, and the effect of large-scale turbulence is considered. Our model, which overcomes many severe challenges to the reprocessing scheme, can well explain the observed variations in NGC 5548, particularly the correlations and lags among the UV/optical bands. In this work, assuming the corona heating is associated with turbulences in the inner accretion disk, we extend our study to model the correlations and lags between the X-ray and UV/optical bands. We find that our model, without the need of light echoing, can well reproduce the observed UV to X-ray lags and the optical to UV lags simultaneously in four local Seyfert galaxies, including NGC 4151, NGC 4395, NGC 4593, and NGC 5548. In our scenario, relatively larger UV to X-ray lag is expected for AGN with smaller innermost disk radius and thus more compact corona. Interestingly, for these Seyfert galaxies studied in this work, sources with relatively larger UV to X-ray lags do have broader Fe K$\alpha$ lines, indicative of relativistic broadening due to more compact corona and smaller innermost disk radius. If confirmed with more X-ray and UV monitoring campaigns, this interesting discovery would provide a new probe to the inner disk/corona.

\end{abstract}


\section{Introduction}\label{sec:intro}

Since unveiled by \citet{LyndenBell1969}, the central engine of active galactic nuclei (AGNs) has been understood through an accreting supermassive black hole (BH), surrounded by an optically thick, geometrically thin accretion disk \citep{ShakuraSunyaev1973}, threaded by strong magnetic fields to form a hot corona \citep{Galeev1979,HaardtMaraschi1991}. The so-called ``big blue bump'' in the UV/optical is thought to be the thermal emission from the thin accretion disk, while the X-ray emission comes from the hot corona above inner regions of the thin disk. 

If the central X-ray emission illuminates the thin disk, it would be absorbed/reflected by the disk, and the absorbed X-ray emission would then be reprocessed and, speculatively, re-emitted thermally. Furthermore, if the UV/optical variation is dominantly driven by the X-ray variation, as assumed in the conventional lamppost reprocessing model \citep[e.g.,][]{Krolik1991}, the implied UV/optical lag--wavelength relation is qualitatively consistent with that observed \citep[e.g.,][]{Fausnaugh2016,Starkey2017}. However, the reprocessing scenario has been challenged in many aspects \citep[cf.][and references therein]{Zhu2018,Cai2018}, including the deficit of energy budget \citep[e.g.,][]{Edelson1996,Gaskell2007} and the timescale-dependent color variation \citep{Sun2014,Zhu2016,Zhu2018,Cai2019}.

A new challenge against the reprocessing scenario has recently emerged thanks to the unique capability of the {\it Swift} satellite, capable of quasi-simultaneously monitoring the X-ray and the UV/optical. For several local Seyfert galaxies, such as NGC 4151 \citep{Edelson2017} and NGC 4593 \citep{McHardy2018}, the observed UV to X-ray lags\footnote{Throughout, a positive UV to X-ray lag means the UV variation lags behind the X-ray variation.} are too large to be afforded by the reprocessing scenario. 
To account for these large lags, either an additional extreme-UV reprocessor, delaying the X-ray heating onto the outer disk \citep{GardnerDone2017}, or a strong contamination by the diffuse large-scale emissions from the broad line region \citep{KoristaGoad2001,McHardy2018,Cackett2018,Chelouche2018,KoristaGoad2019,Mahmoud2019} have been proposed, both demanding significant modification to the standard thin disk scheme.

Persisting in the classic thin disk diagram, we previously developed a model, exploring the UV/optical continuum lag in AGNs (EUCLIA), in which the UV/optical variations are attributed to disk turbulences, but not the light echoing \citep{Cai2018}. Our model, which overcomes many severe challenges to the reprocessing scheme, can well explain the observed variations in NGC 5548, particularly correlations, lags, and the timescale-dependent color variations among the UV/optical bands. Owing to quicker regression capability of local fluctuations at smaller radii when responding to the large-scale turbulence, the emission of longer wavelength coming from the outer disk regions would lag that of the shorter wavelength from the inner disk regions.

Assuming the corona heating is also associated with turbulences in the inner accretion disk \citep[e.g.,][]{Kang2018}, in this work, we extend our study to model the variation of coronal X-ray emission as laid out in Section~\ref{sec:model}. In Section~\ref{sect:results}, we find that our model, without the need of light echoing, can well reproduce the observed UV to X-ray lags in four local Seyfert galaxies, covering a broad range of BH mass introduced in Section~\ref{sec:sample}. In our scenario, a relatively larger UV to X-ray lag is expected for AGN with a smaller innermost accretion disk radius and thus more compact corona. Interestingly, as discussed in Section~\ref{sect:discussion}, for these Seyfert galaxies studied in this work, sources with relatively larger UV to X-ray lags do have more relativistically broadened Fe K$\alpha$ lines, indicative of more compact corona and smaller innermost disk radius. Finally, conclusions are presented in Section~\ref{sect:conclusion}.

\section{Model ingredients}\label{sec:model}

Following \citet{Cai2016} we have split the cold thin disk into square-like zones in $r$ and $\phi$ directions from an inner edge, $r_{\rm in} \geqslant r_{\rm ISCO}$, to a large enough outer one, $r_{\rm out} = r_{\rm in} f^{N_{\rm r}}_{\rm rbr} \gtrsim 5000~r_{\rm g}$, where $r_{\rm ISCO}$ is the spin-dependent innermost stable circular orbit (ISCO) radius, $N_{\rm r}$ is the number of layers, $f_{\rm rbr} = 1.1$ is the radial boundary ratio of each layer, and $r_{\rm g} \equiv G M_\bullet / c^2$ is the gravitational radius. We assume the Novikov-Thorne relativistic effective temperature profile, $T_{\rm NT}(r|M_\bullet,~\dot M_{\rm acc},~a_*)$, as a function of radius, $r$, for BH mass $M_\bullet$, accretion rate $\dot M_{\rm acc}$ (or Eddington ratio\footnote{Here, $\lambda_{\rm Edd} \equiv L_{\rm bol}/L_{\rm Edd}(M_\bullet)$  and $\dot M_{\rm acc} = \lambda_{\rm Edd} L_{\rm Edd}(M_\bullet) / \eta c^2$, where $L_{\rm bol}$ and $L_{\rm Edd}$ is the bolometric and Eddington luminosities, respectively,  and $\eta$ is a spin-dependent efficiency (e.g., $\eta = 0.23$ for $a_* = 0.98$; $\eta = 0.057$ for $a_* = 0$).}
$\lambda_{\rm Edd}$), and BH spin $a_*$ \citep{NovikovThorne1973}. 

Stimulated by the independently fluctuating inhomogeneous disk model of \citet{DexterAgol2011}, \citet{Cai2016} propose that a revised inhomogeneous disk model can well reproduce the timescale-dependent color variation if the characteristic timescale of thermal fluctuation in the disk is radius-dependent. In this upgraded model, the local temperature fluctuates around the mean, $\log T_{\rm return} = \log T_{\rm NT} - 2 \sigma^2_{\rm l,indep}\ln(10)$, and its fluctuation possibly driven by magnetic turbulence is described by the first-order continuous autoregressive process \citep{Kelly2009} with a radius-dependent damping timescale of $\tau(r) = \tau_0 (r/r_{\rm g})^\alpha$ and a radius-independent long-term variance of $\sigma^2_{\rm l,indep}$, or, equivalently, a short-term variance of $\sigma^2_{\rm s,indep}(r) = 2 \sigma^2_{\rm l,indep}/\tau(r)$.

Furthermore, to account for the correlations and lags among the UV/optical continua, \citet{Cai2018} propose that the local temperature fluctuation is coupled with a common large-scale fluctuation as a net effect of many distinct outward/inward propagations. This common large-scale fluctuation is depicted with a short-term variation amplitude, $\sigma_{\rm s,com}(r) = f_{\rm com/indep} \times \sigma_{\rm s,indep}(r) \times (r/r_{\rm g})^{\gamma+\alpha/2}$, complementing with a radius-independent smoothing timescale, $\tau_{\rm cut,0}$, where $f_{\rm com/indep} \equiv \sigma_{\rm s,com}(r_{\rm g}) / \sigma_{\rm s,indep}(r_{\rm g})$ is the relative contribution between large-scale and local fluctuations at $r_{\rm g}$ (cf. \citealt{Cai2018} for details). 

In \citet{Cai2018}, we have explored those aforementioned parameters, namely $\tau_0 \equiv \tau(r_{\rm g}) = 1/6$ day, $\alpha = 1$, $\sigma_{\rm l,indep} = 0.08$ dex, $f_{\rm com/indep} = 1.5 \sqrt{6}$, $\gamma = -1$, and $\tau_{\rm cut,0} = 10$ day, for NGC 5548 with $M_\bullet = 5\times 10^7 M_\sun$. In this work, considering four sources with different BH masses, we have assumed in our baseline model that both the damping and smoothing timescales are simply scaling with $M_\bullet$ as $\tau(r, M_\bullet) = \tau_0 ({r}/{r_{\rm g}})^\alpha ({M_\bullet}/{5\times 10^7 M_\sun})$ day and $\tau_{\rm cut}(M_\bullet) = \tau_{\rm cut,0} ({M_\bullet}/{5\times 10^7 M_\sun})$ day, respectively, while the remaining parameters are mass-independent. We have adopted the same parameters as those previously explored for NGC 5548, except $\sigma_{\rm l,indep}$ and $f_{\rm com/indep}$, since we find that the correlations and lags between the X-ray and UV/optical bands for the four sources can be quite well simultaneously modeled using $\sigma_{\rm l,indep} = 0.2$ dex and $f_{\rm com/indep} = 1.5$ (see Section~\ref{sect:results}).

To approximate the radiative transfer through the disk atmosphere, the cold thin disk emission is assumed to be a color-temperature corrected blackbody, i.e., $B^{\rm col}_\nu[\nu, T_{\rm NT}] = B_\nu[\nu, f_{\rm col} T_{\rm NT}] / f^4_{\rm col}$, where $\nu$ is the photon frequency, $B_\nu$ is the blackbody radiation intensity, and $f_{\rm col}$ is the color-temperature correction following \citet{Done2012}.

A major improvement of this work is introducing a classical treatment of X-ray emission for the hot corona, located above the inner cold disk region from $r_{\rm in}$ to $r_{\rm corona}$\footnote{
Below $r_{\rm in}$, the accretion flow becomes radiatively inefficient since most of the power may be released via winds/jets \citep{YuanNarayan2014,Hall2018,Sun2019a}.}. A radius-independent fraction, $f_{\rm corona}$, of the released energy is assumed to be dissipated to the corona, and then re-emitted as a power law spectrum of $P_\nu(\nu) \sim \nu^{1-\Gamma} \exp(- h \nu /E_{\rm max,cut} - E_{\rm min,cut}/h\nu)$, characterized by a photon index, $\Gamma$, and high (low) energy cutoff, $E_{\rm max,cut}~(E_{\rm min,cut})$. 
The normalization of $P_\nu(\nu)$ is then constrained by the energy conservation of each zone and is given by $\int_\nu P_\nu d\nu = f_{\rm corona} \int_\nu B^{\rm col}_\nu d\nu$. In this work, without affecting our conclusions, we simply assume $\Gamma = 1.9$, $f_{\rm corona} = 0.5$, $E_{\rm min,cut} = 0.1$ keV, and $E_{\rm max,cut} = 100$ keV \citep{BrandtAlexander2015,Molina2019}.

Therefore, if the disk at redshift $z$ is viewed with an inclination angle $i$, the flux density observed at $\nu_{\rm obs}$ per frequency per area is given by
\begin{eqnarray}
	 &\tilde F_{\nu}(\nu_{\rm obs}, z, t) = (1 + z) \frac{\cos i}{\pi D^2_{\rm L}(z)} \times \int^{2 \pi}_{\phi=0}  d\phi \nonumber \\
	 &\Big\{ \int^{r_{\rm corona}}_{r_{\rm in}} \pi [ (1 - f_{\rm corona}) B^{\rm col}_\nu(\nu_{\rm e}, \tilde T_{\rm NT}) + \tilde P_\nu(\nu_{\rm e}, r, t)] r dr  \nonumber \\
	 &+ \int^{r_{\rm out}}_{r_{\rm corona}} \pi B^{\rm col}_\nu[\nu_{\rm e}, \tilde T_{\rm NT}(r, t)] r dr \Big\},
\end{eqnarray}
where fluctuating quantities are indicated with a tilde over them, $D_{\rm L}(z)$ is the luminosity distance, and $\nu_{\rm e} = \nu_{\rm obs} (1+z)$. We take $i = 45^\circ$ as a reference.

As illustrated above, a major assumption in the model extension is that,  for any individual inner disk zone within $r_{\rm in}$ -- $r_{\rm corona}$, 
the X-ray corona emission it contributes has the same variation pattern with the UV/optical emission, i.e., both modulated by an identical local independent fluctuation and the common large-scale fluctuation.
This assumption is generally supported by the discovery of \citet{Kang2018} that the corona heating is closely associated with magnetic turbulence \citep[see also, e.g.,][from a theoretical point of view]{LiuMineshigeShibata2002}.
In fact, the essence of the assumption is that, the fluctuation of X-ray emission is similarly modulated by the common large-scale fluctuation
(see \citealt{Cai2018} for possible mechanisms responsible for the common large-scale fluctuation).
An identical local corona and disk fluctuation in each zone is actually not a necessary condition.  For example, in a ``patch'' scheme, we could allow a certain fraction, say $f_{\rm corona}$, 
of the individual inner disk zones within $r_{\rm in}$ -- $r_{\rm corona}$ to dissipate 100\% of the energy into the corona while others contribute zero. 
In this way, the modeled local corona fluctuation is totally independent to that of the disk. 
Such a modification however does not affect any of the analyses in this work.
This is because the X-ray emission region barely overlaps with the disk regions that dominate the observed UV/optical emission production, thus their local fluctuations are already independent to each other.
Being numerically confirmed, such a modification for this ``patch'' scheme only results in slightly smaller correlations and subtly larger scatter of lags between the X-ray and UV/optical bands. To weaken the assumption of an identical local corona and disk fluctuation in each zone, we should present our results within this ``patch'' scheme in the following.

Another key point of the assumption is that the characteristic timescale of the corona fluctuation is the same as that of the underlying disk, i.e., proportional to $r$. This is expected if the corona and
the innermost disk fluctuations are both driven by magnetic turbulences.  

The illumination of the X-ray emission onto the underlying disk zone could also play a role in yielding coupled variation between the corona and disk emission. 
However, in the current model we ignore the illumination process (cf. Section~\ref{sec:intro} and \citealt{Cai2018}), and leave the modeling of both disk fluctuation and X-ray reprocessing to a future dedicated work.
In this work, we focus on whether the fluctuation model alone can explain the observed lags between the X-ray and UV/optical bands for which the reprocessing model failed to account.

\begin{figure*}[t!]
\centering
\includegraphics[width=\textwidth]{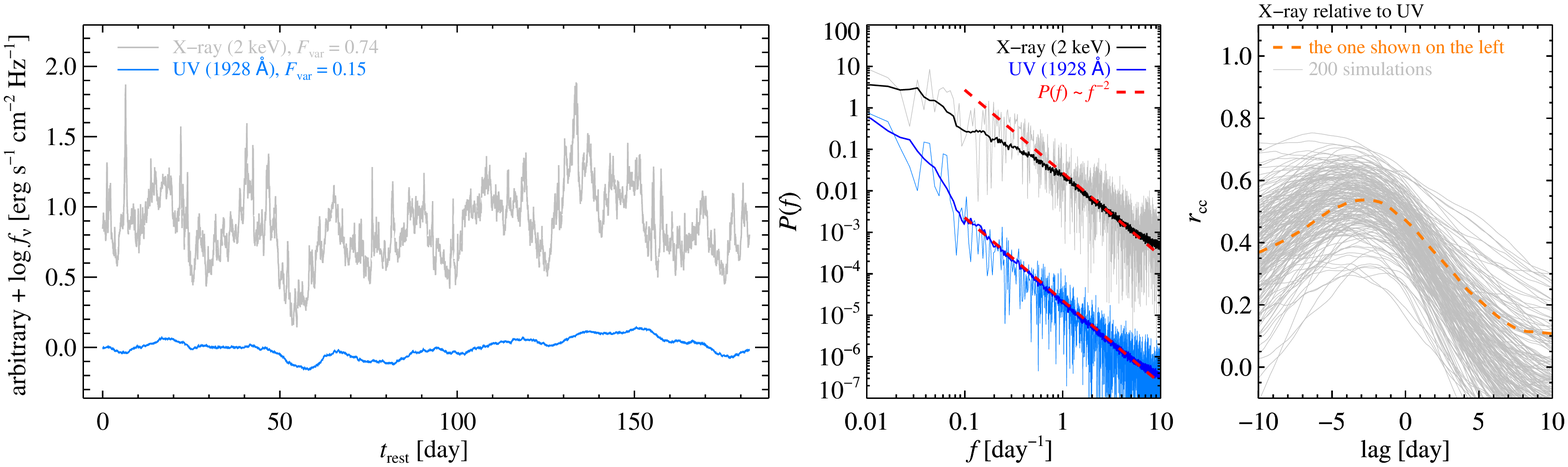}
\caption{
Left panel: an example of the simulated X-ray (2 keV; light-gray line) and UV (1928~\angstrom; light-blue line) light curves for $M_\bullet = 5 \times 10^7~M_\odot$, $\lambda_{\rm Edd} = 0.05$, $a_* = 0$, $r_{\rm in} = 6~r_{\rm g}$, $r_{\rm corona} \simeq 2~r_{\rm in}$, and an time length of $\sim 180$ days. 
Middle panel: the highly unsteady light-gray (X-ray) and light-blue (UV) lines for the power density spectra corresponding to the light curves in the same colors on the left panel, while the relatively smooth black (X-ray) and blue (UV) lines for the median power density spectra of 200 ideal simulations, compared to the power density spectrum of $\sim f^{-2}$ (red dashed lines). 
Right panel: the cross-correlation function of X-ray relative to UV for the light curves shown on the left panel (orange dashed line) versus those for each of the 200 simulations (light-gray lines).
}\label{fig:lc_ccf_psd}
\end{figure*}

\begin{figure*}[t!]
\centering
\includegraphics[width=\textwidth]{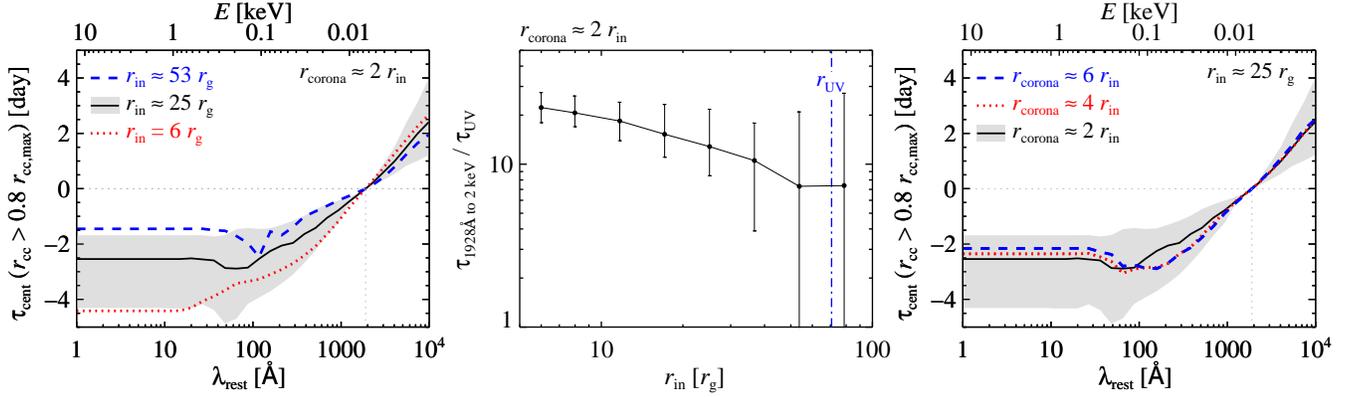}
\caption{
Left panel: the rest-frame lag-wavelength relations for the same $M_\bullet = 5 \times 10^7~M_\odot$, $\lambda_{\rm Edd} = 0.05$, $a_* = 0$, and $r_{\rm corona} \simeq 2~r_{\rm in}$, but different $r_{\rm in} = 6$, $\simeq 25$, and $\simeq 53~r_{\rm g}$. Lines are the medians of 200 simulations for different $r_{\rm in}$ while the light-gray region is the 25-75$^{\rm th}$ percentile ranges only for $r_{\rm in} \simeq 25~r_{\rm g}$. 
Middle panel: the median (filled circles) and 25-75$^{\rm th}$ percentile ranges (vertical bars) of the UV (1928~\angstrom) to X-ray (2 keV) lags inferred from simulations with different $r_{\rm in}$ are normalized by the UV light-crossing timescale, $\tau_{\rm UV} \equiv r_{\rm UV}/c$, where $r_{\rm UV}$ is the flux-weighted UV-emitting radius (the dot-dashed line; see Section~\ref{sec:diff_rin}). 
Right panel: same as the left panel, but for fixed $r_{\rm in} \simeq 25~r_{\rm g}$ and different $r_{\rm corona} \simeq 2$, 4, and $6~r_{\rm in}$.
}\label{fig:diff_r_in}
\end{figure*}

\subsection{Simulation approach and basic properties of ideal light curves}\label{sect:sim_app}

In fine steps of $\Delta t_{\rm rest}$ in the rest-frame (see Section~\ref{sec:sample} and Table~\ref{tab:model_par}), we firstly simulate the fluctuating spectral energy distributions (SEDs) from the X-ray to the UV/optical, from which ideal monochromatic light curves are extracted at 40 wavelengths sampled evenly in logarithm from 1~\angstrom~to 1 $\mu$m, besides two specific wavelengths at 2 keV (a typical X-ray wavelength) and 1928~\angstrom~(the pivot wavelength of the {\it Swift}-{\it UVW}2 band). Following Equation~2 of \citet[][see also, e.g., \citealt{Edelson1990,Rodriguez-Pascual1997,Vaughan2003}]{Fausnaugh2016}, the fractional root-mean-square variability amplitude is defined by $F_{\rm var} \equiv \sqrt{\sum^{N}_{i=1} [\tilde F_\nu(t_i) / \langle \tilde F_\nu \rangle_t - 1]^2/N}$, where $\tilde F_\nu(t_i)$ is the value of the light curve at epoch $t_i$, $\langle \tilde F_\nu \rangle_t$ is the mean value of the light curve, and $N$ is the total number of epochs. The maximal correlation coefficient, $r_{\rm cc,max}$, and the centroid lag averaging correlation coefficients larger than 80\% of the maximal one, $\tau_{\rm cent}$, are obtained relative to the UV band (1928~\angstrom) without detrending, by performing the linearly interpolated cross-correlation analysis \citep{Peterson1998}. Totally, we repeat 200 times of independent simulations.

Figure~\ref{fig:lc_ccf_psd} shows an example of the simulated X-ray and UV light curves, where a clear correlation between them is intuitive. Compared to the UV, larger $F_{\rm var}$ and more high-frequency power are found in the X-ray, as often seen in AGNs. Averaging 200 simulations, the high-frequency power density spectra for both the UV and the X-ray are similar to that of $\sim f^{-2}$, but there is a break at low-frequency for the X-ray, qualitatively consistent with that found by \citet{Czerny1999}. 
A more quantitative comparison is deferred to Section~\ref{sect:ngc5548}.

The lag of the X-ray relative to the UV is generally negative, indicating the UV variation lags behind the X-ray. Interestingly, the exact amount of the lag may change from one monitoring period to another with the same cadence (see Section~\ref{sect:randomness} for further discussion).

\subsection{Increasing the UV to X-ray lag with decreasing the inner edge of the cold accretion disk}\label{sec:diff_rin}

Comparing the different amount of the UV to X-ray lags between NGC 4151 and NGC 5548, which have similar BH mass, \citet{Edelson2017} propose that an extreme-UV torus could enlarge the UV to X-ray lag for NGC 4151, but speculate that the smaller UV to X-ray lag for NGC 5548 is probably due to the lower single-to-noise ratio of its X-ray data.

Instead, we physically attribute the amount of the UV to X-ray lag to the different inner edge of the cold accretion disk or the compactness of corona. This point is clearly illustrated in Figure~\ref{fig:diff_r_in}. If the inner edge of the cold disk is smaller, the UV to X-ray lag increases. Note that in our model the characteristic timescale of local fluctuation is smaller at smaller radius; the UV to X-ray lag increases with increasing the timescale difference of temperature fluctuations between regions emitting the X-ray and the UV. Therefore, once the disk extends more inward, the more inner disk regions with smaller timescales of temperature fluctuation would be involved in producing the X-ray emission, and then, the UV to X-ray lag would increase with decreasing the inner disk radius. 

Conversely, the UV to X-ray lag decreases with increasing the outer edge of the hot corona (cf. the right panel of Figure~\ref{fig:diff_r_in}), but the effect is negligible due to weaker X-ray contribution from the outer disk regions. Considering this relative insensitivity of the UV to X-ray lag on $r_{\rm corona}$, we fix $r_{\rm corona} \simeq 2~r_{\rm in}$ in the following analysis.

Normalized by the UV light-crossing timescale, $\tau_{\rm UV} \equiv r_{\rm UV}/c$, where $r_{\rm UV} = r_{\rm UV}$(1928~\angstrom) is the flux-weighted UV-emitting radius of the standard thin disk \citep[cf. Equation 10 of ][]{Fausnaugh2016}, the middle panel of Figure~\ref{fig:diff_r_in} directly illustrates the UV to X-ray lag in our model could be larger than that of the reprocessing model by a factor of up to $\sim 20$, increasing with decreasing $r_{\rm in}$. Note that the scatter of the UV to X-ray lag increases with increasing $r_{\rm in}$ and becomes interestingly large when $r_{\rm in}$ is comparable to $r_{\rm UV}$ (see Section~\ref{sect:randomness} for further discussion).

\begin{deluxetable*}{lcc ccc ccc}[t!]
\tablenum{1}
\tablecaption{Basic parameters in the baseline model and observational information for those concerned Seyfert Galaxies\label{tab:model_par}}
\tablewidth{0pt}
\tablehead{
\colhead{} & \colhead{Redshift} & \colhead{BH Mass} & \colhead{Eddington} & \colhead{Spin} & \colhead{Inner Most Radius of} & \colhead{Campaign Duration} & \colhead{Median Sampling} & \colhead{Simulated Time} \\
\colhead{Object} & \colhead{$z$} & \colhead{$M_\bullet/[10^6~M_\odot]$} & \colhead{Ratio $\lambda_{\rm Edd}$} & \colhead{$a_*$} & \colhead{Cold Disk $r_{\rm in}/[r_{\rm g}]$} & \colhead{${\cal CD}_{\rm obs}/$[days]} & \colhead{Interval [days]} & \colhead{Step $\Delta t_{\rm rest}/$[days]} \\
\colhead{(1)} & \colhead{(2)} & \colhead{(3)} & \colhead{(4)} & \colhead{(5)} & \colhead{(6)} & \colhead{(7)} & \colhead{(8)} & \colhead{(9)}
}
\startdata
NGC 4151 & 0.00326 &  40     & 0.02   & 0.98  & 1.61                  & $69.3$    & 0.22   & 0.02 \\
NGC 4593 & 0.00834 &  7.5    & 0.08   & 0     & 6                     & $22.6$    & 0.12   & 0.01 \\
NGC 5548 & 0.01627 &  50     & 0.05   & 0     & $\sim 36.7^\dagger$     & $175.3$   & 0.44   & 0.04 \\
NGC 4395 & 0.00106 &  0.01   & 0.04   & 0     & $\sim 53.7^\dagger$     & $0.6$    & 0.0023 & 0.0002 \\
\enddata
\tablecomments{\footnotesize 
Column 1: The name of object. 
Column 2: The redshift from the SIMBAD database (http://simbad.u-strasbg.fr/simbad/). 
Column 3: The BH mass consistent with \citet{BentzKatz2015}, except NGC 4395 which is from \citet{Woo2019}. 
Column 4: The Eddington ratio $\lambda_{\rm Edd} \equiv L_{\rm bol}/L_{\rm Edd}$ consistent with \citet{McHardy2018}, except NGC 4395 for which we have re-derived its new Eddington ratio with $L_{\rm bol}$ from \citet{McHardy2018} given the new BH mass. 
Column 5: Constraint on the BH spin for NGC 4151 is given by \citet{Keck2015}, while spins of the other sources without robust constraint are assumed to zero. 
Column 6: The adopted inner most radius of the cold disk in our baseline model for each source. The value with dagger is the only one parameter subjectively selected according to the observed UV to X-ray lags.
Columns 7 and 8: The observed-frame campaign duration and median sampling interval for sources monitored by {\it Swift} (NGC 4151: \citealt{Edelson2017}; NGC 4593: \citealt{McHardy2018}; NGC 5548: \citealt{Edelson2015}) and {\it XMM-Newton} (NGC 4395: \citealt{McHardy2016}).
Column 9: The rest-frame time step adopted in our simulation, better than the real sampling interval for each source by about a factor of ten.
}
\end{deluxetable*}

\section{Modeling local Seyfert galaxies}\label{sec:sample}

To concisely illustrate that our model is capable of explaining the puzzling large UV to X-ray lags, we focus on four local Seyfert galaxies, discussed by \citet{McHardy2018}, i.e., NGC 4151 \citep{Edelson2017}, NGC 4593 \citep{McHardy2018}, NGC 5548 \citep{Edelson2015,Fausnaugh2016}, and NGC 4395 \citep{Cameron2012,McHardy2016}, covering a broad range of BH mass and $\sim 1\%-8\%$ of Eddington ratio. NGC 2617 \citep{Shappee2014} is currently excluded due to its large error on the derived UV to X-ray lag.

The most relevant quantities for these sources are tabulated in Table~\ref{tab:model_par}. The $M_\bullet$ and $\lambda_{\rm Edd}$ are consistent with those quoted by \citet{McHardy2018}, except NGC 4395 for which we adopt the newest BH mass \citep{Woo2019} and re-derive the corresponding Eddington ratio.

Measuring the spin of supermassive BH is important but still challenging when using the relativistic X-ray reflection method \citep[e.g.,][]{Reynolds2014,Kammoun2018}.
Although most of these concerned Seyfert galaxies have been extensively studied in X-ray \citep[e.g.,][]{Nandra1997c,Nandra2007,ShuXW2010b,Walton2013}, up to now, only the BH spin of NGC 4151 has been found to be near-maximal \citep{Keck2015,Beuchert2017}, while ambiguous spins are still assumed in literature for the other sources (e.g., NGC 4593: \citealt{Ursini2016a}; NGC 5548: \citealt{Kammoun2019a}; NGC 4395: \citealt{Kammoun2019b}). Therefore, we assume zero spin for the later three sources.
As demonstrated in Section~\ref{sec:diff_rin}, the UV to X-ray lag would increase with increasing spin if $r_{\rm in} = r_{\rm ISCO}(a_*)$.
However, if $r_{\rm in} $ is significantly larger than $ r_{\rm ISCO}$, as respectively discussed for NGC 5548 and NGC 4395 in Sections~\ref{sect:ngc5548} and \ref{sect:ngc4395}, the uncertain spin would have a negligible effect on the UV to X-ray lag.

\subsection{Re-estimation of observed cross-correlation properties}\label{sect:rs_ccf}

We intend to compare our model to observations on both $r_{\rm cc,max}$ and $\tau_{\rm cent}$ measured relative to the {\it Swift}-{\rm UVW2} band, but not all of them were available in literature and $r_{\rm cc,max}$ values are generally reported without errors. Therefore, we take the observed light curves from literature for these sources (unavailable for NGC 4395) to re-estimate their interband correlations and lags, adopting the linearly interpolated cross-correlation function method\footnote{PyCCF: \url{https://ascl.net/code/v/1868}} \citep[ICCF; e.g.,][]{GaskellPeterson1987,Peterson1998,Peterson2004,Sun2019a,Edelson2019}. The {\it Swift}-{\rm UVW2} band is always taken as reference, relative to which the correlations and lags of the other bands are measured, implemented the ``2-way'' interpolation \citep[see, e.g.,][]{Edelson2019}. Their 1$\sigma$ uncertainties are estimated as the standard deviations of $10^3$ realizations generated using the standard flux randomization/random subset selection method. Note that as discussed in \citet{Cai2018} no detrending has been applied. 
Our measurements of $r_{\rm cc,max}$ and $\tau_{\rm cent}$ are generally consistent with those published when available, and in the following we compare our model with both our own measurements from observations and those found in literature.

\subsection{Simulating ``real" light curves}\label{sect:real_sim}

In Section~\ref{sect:sim_app}, we have introduced simulating the ideal monochromatic light curves for demonstrating their basic properties implied by our model.
Furthermore, to faithfully compare with observations,
we convolve our redshifted fluctuating SEDs with the {\it Swift} six UVOT transmission curves and the corresponding specific X-ray bandwidths of {\it Swift} as reported in the relevant literatures. For NGC 5548, we consider more UV/optical bands following \citet{Fausnaugh2016}.

For each source, ideal broad-band light curves are generated with an observed-frame time step of $\Delta t_{\rm rest} (1+z)$ better than, and an appropriate time length of ${\cal CD}_{\rm obs}$ comparable to, those of its real monitoring campaign, respectively (see Table~\ref{tab:model_par}).
For each band, further taking into account the real sampling and randomly fluctuating fluxes according to the observed photometric error of each epoch, we obtain the mock ``real'' light curve \citep[cf. Section 3.2 in][]{Cai2018}.
Then, the interband cross-correlation properties are similarly derived as described in Section~\ref{sect:rs_ccf}.
Note that the host galaxy contamination has not been added to these ``real'' light curves, as our analyses of correlation functions are not affect by the constant host emission. 

In analog to the simulated ideal monochromatic light curves presented in Section~\ref{sect:sim_app}, we also obtain 200 simulated ``real" light curves for each source. To statistically compare with observations on the cross-correlation properties, i.e., $r_{\rm cc,max}$ and $\tau_{\rm cent}$, we present their median, 25-75$^{\rm th}$, and 10-90$^{\rm th}$ percentile ranges inferred from these simulations.

\section{Results}\label{sect:results}

\subsection{The puzzling large UV to X-ray lags: NGC 4151 and NGC 4593}\label{sect:ngc4151_ngc4593}

\begin{figure*}[t!]
\centering
\includegraphics[width=0.4\textwidth]{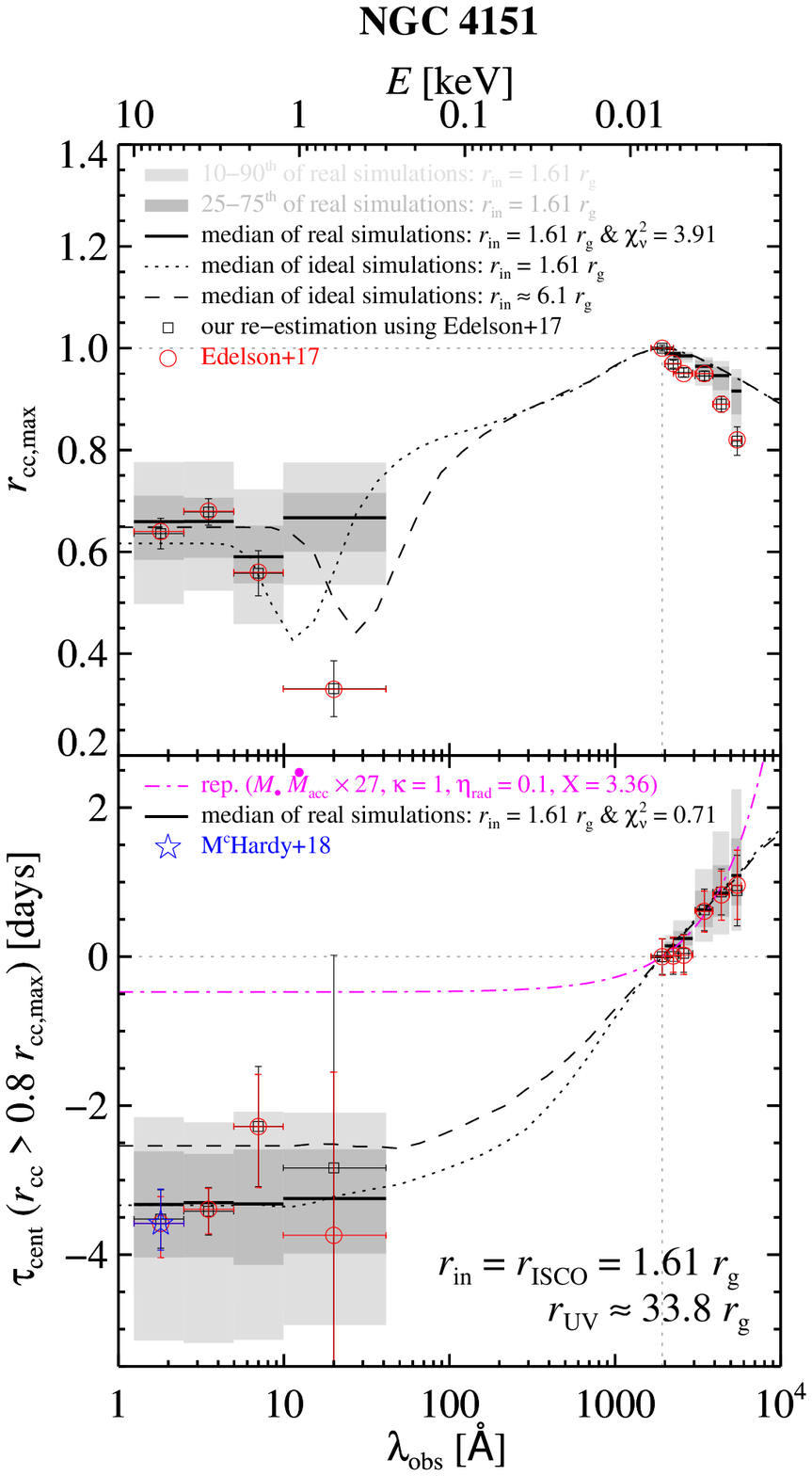}
\includegraphics[width=0.4\textwidth]{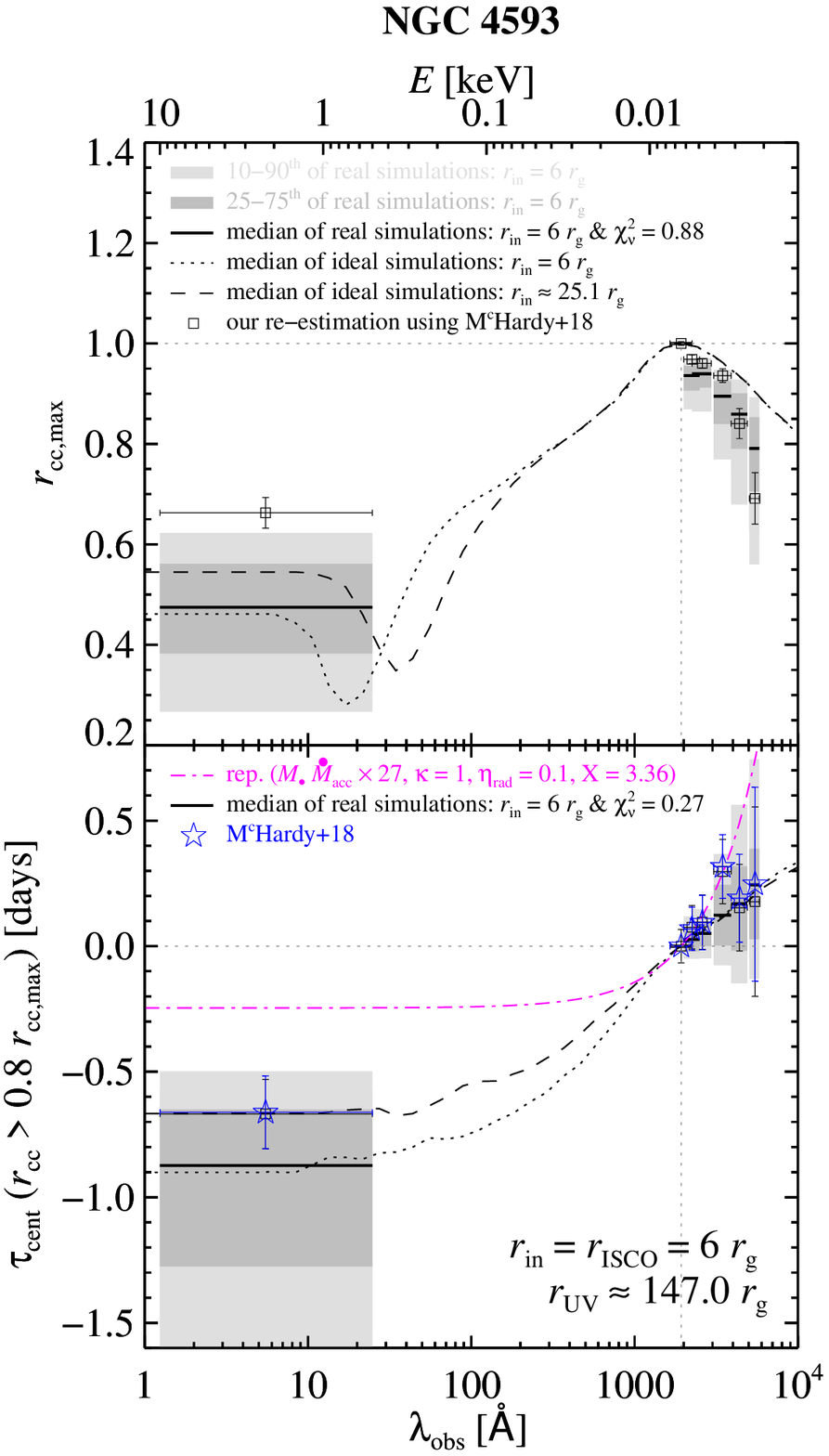}
\caption{ 
Relative to the {\it Swift}-{\it UVW2} band (1928~\angstrom), the maximal correlation coefficient, $r_{\rm cc,max}$ (top panels), and the centroid lag, $\tau_{\rm cent}$ (bottom panels), as a function of the observed wavelength for NGC 4151 (left panels) and NGC 4593 (right panels), respectively. 
The observational data are from \citet[][red open circles]{Edelson2017} and \citet[][blue open stars]{McHardy2018} as well as our own measurements (black open squares; see Section~\ref{sect:rs_ccf}).
In each panel, illustrated are the median (black thick solid horizontal lines), 25-75$^{\rm th}$ (gray regions), and 10-90$^{\rm th}$ (light-gray regions) percentile ranges of 200 ``real'' simulations (see Section~\ref{sect:real_sim}) as well as the median (black dotted lines) of 200 ideal simulations (see Section~\ref{sect:sim_app}) for our baseline model with $r_{\rm in} = r_{\rm ISCO}$ (i.e., $1.61~r_{\rm g}$ for NGC 4151 and $6~r_{\rm g}$ for NGC 4593).
The nominated $\chi^2_\nu$ quantitatively assesses the difference between our ``real'' simulations and observations (see Section~\ref{sect:ngc4151_ngc4593}).
Shown for comparison are the median results (black dashed lines) of 200 ideal simulations with $r_{\rm in} > r_{\rm ISCO}$ (i.e., $r_{\rm in} \simeq 6.1~r_{\rm g}$ for NGC 4151 and $r_{\rm in} \simeq 25.1~r_{\rm g}$ for NGC 4593) as well as the lag-wavelength relation implied by the reprocessing model, where $M_\bullet \dot M_{\rm acc}$ has been increased by a factor of 27 or equivalently lag by a factor of 3 (magenta dot-dashed lines). 
The fiducial $r_{\rm in}$ and $r_{\rm UV}$(1928~\angstrom) for each source are also nominated in the lower-right corner.
These two galaxies are grouped owing to their large UV to X-ray lags, compared to that implied by the reprocessing model.
}\label{fig:large_UV2Xray}
\end{figure*}

\begin{figure}[t!]
\centering
\includegraphics[width=0.4\textwidth]{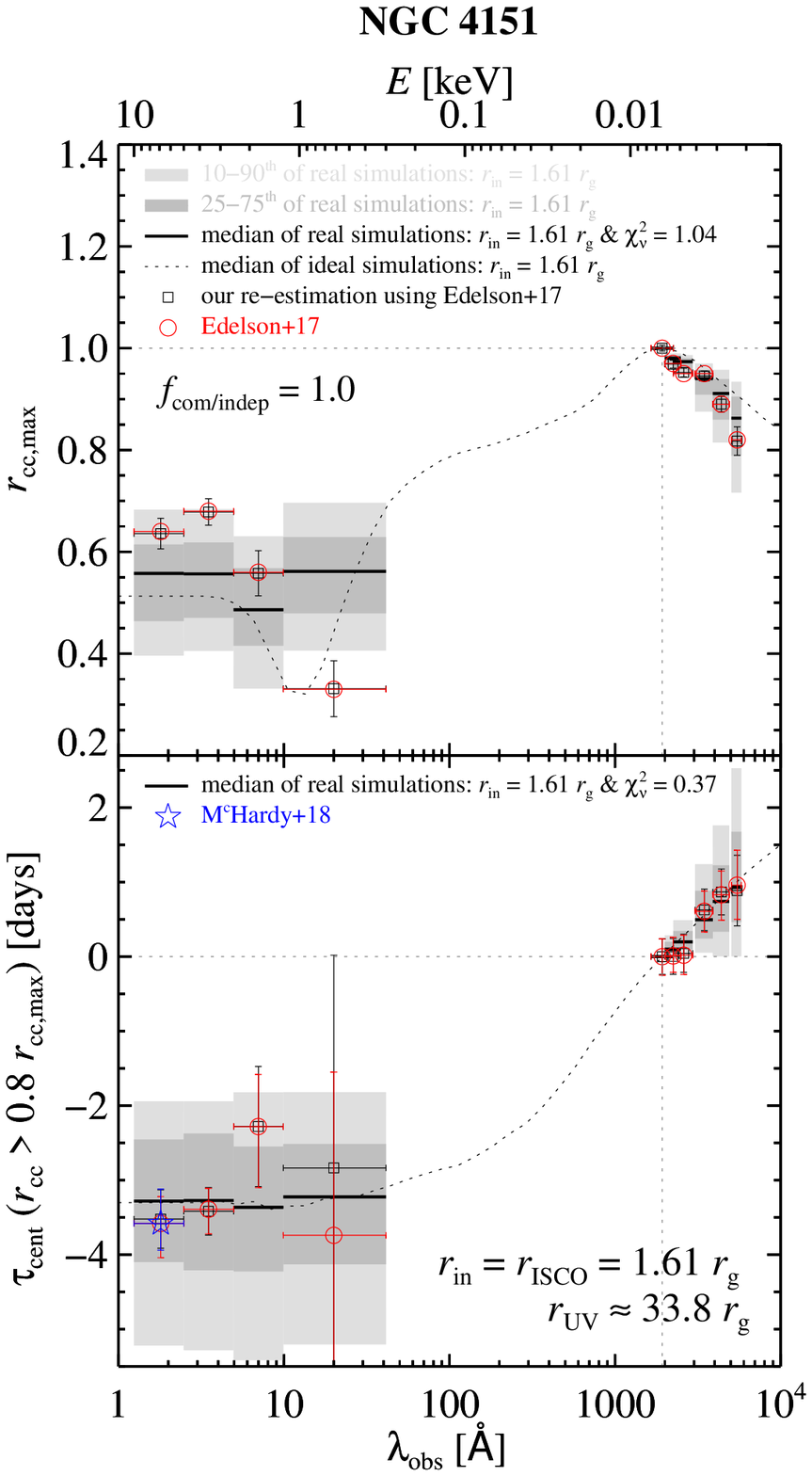}
\caption{In analog to the left panels of Figure~\ref{fig:large_UV2Xray} for NGC 4151, but for a model with different $f_{\rm com/indep}$. 
Here the illustrated model has the same $r_{\rm in}$ as but smaller $f_{\rm com/indep}~(= 1.0)$ than the baseline model with $f_{\rm com/indep} = 1.5$ presented in the left panels of Figure~\ref{fig:large_UV2Xray}.
}\label{fig:ngc4151_better}
\end{figure}

As introduced in Section~\ref{sect:rs_ccf}, we have re-estimated the correlations, $r_{\rm cc,max}$, and lags, $\tau_{\rm cent}$, using the observed {\it Swift} light curves, and our measurements are
generally consistent with those available in literature. 
Relative to the UV band, i.e., {\it Swift}-{\it UVW2}, Figure~\ref{fig:large_UV2Xray} shows our measurements (along with those from literature) of $r_{\rm cc,max}$ (top panels) and $\tau_{\rm cent}$ (bottom panels) as a function of the observed wavelength for NGC 4151 (left panels) and NGC 4593 (right panels).

The UV to X-ray lags of these two galaxies have recently reported to be significantly larger than that implied by the reprocessing model \citep{Edelson2017,McHardy2018}.
This new challenge against the reprocessing model is also illustrated in the lower panels of Figure~\ref{fig:large_UV2Xray}, where shown are the lag-wavelength relations implied by the reprocessing model (the magenta dot-dashed lines), using Equation (12) of \citet[]{Fausnaugh2016} and assuming the radiation efficiency $\eta_{\rm rad} = 0.1$, the ratio of external to internal heating, $\kappa = 1$, the multiplicative scaling factor, $X = 5.04^{3/4}$ \citep{TieKochanek2018a}, and 27 times larger $M_\bullet \dot M_{\rm acc}$ \citep{Zhu2018}.
Note here $M_\bullet \dot M_{\rm acc}$ has to be enlarged by a factor of 27 in order to match the observed UV/optical lags (relative to {\it Swift}-{\it UVW2}).
Even though, the UV to X-ray lags implied by the reprocessing model still appear too small and significantly disagree with the observations for NGC 4151 and NGC 4593. 

In Figures~\ref{fig:large_UV2Xray} we also plot our model predictions for NGC 4151 and NGC 4593 with default parameters (here, $r_{\rm in} = r_{\rm ISCO}$), using both simulated ideal light curves introduced in Section~\ref{sect:sim_app} and ``real" light curves considering the monitoring sampling and flux uncertainties introduced in Section~\ref{sect:real_sim}.
The black dotted lines are the medians of 200 ideal simulations, while the black thick solid horizontal lines, the gray regions, and the light-gray regions are the median, 25-75$^{\rm th}$, and 10-90$^{\rm th}$ percentile ranges of 200 ``real" light curves, respectively.
Note an important prediction of our model is that the interband correlations and lags may change from one simulated light curve to another due to the randomness of the turbulence, and
such scatters are clearly visible with the plotted 25-75$^{\rm th}$, and 10-90$^{\rm th}$ percentile ranges.
After having considered the sampling and flux uncertainties, the medians for $r_{\rm cc,max}$ decreases slightly, comparing with those from the ideal light curves. This is as expected as photometric errors were added to ``real" light curves. The medians for $\tau_{\rm cent}$ are more or less the same, thanks to the high cadence of {\it Swift} monitoring and the small flux uncertainty of $\sim 2\%$. 

It is remarkable that our simulations, without fine tuning any of the default model parameters, appear nicely consistent with the overall trend of the observations. We quantitatively assess the differences between our model and observations using $\chi^2_\nu(x) \equiv \frac{1}{N} \sum^{N}_{i=1} [(x^{\rm m}_i - x_i)/\sigma^{\rm m}_i]^2$, where $x_i$ is the observed value (our own measurements) for $r_{\rm cc,max}$ or $\tau_{\rm cent}$ in the $i$th band, $x^{\rm m}_i$ and $\sigma^{\rm m}_i$ are respectively the corresponding median value and 1$\sigma$ scatter implied by simulations, and $N$ is the number of bands.
Note that to access $\chi^2_\nu$ the {\it Swift}-{\it U} band (and the SDSS-$u$ band of NGC 5548) has been excluded owing to the prominent contamination of emissions from the broad line region \citep{Fausnaugh2016,Edelson2019}.
The derived $\chi^2_\nu(x)$ are presented in Figure~\ref{fig:large_UV2Xray}.
Globally, a well consistence between our model and observations is indicated by the small $\chi^2_\nu(x)$.
In the top-left panel of Figure~\ref{fig:large_UV2Xray}, a noticeable stronger correlation predicted by the baseline model for NGC 4151 can be found in the UV/optical bands. Since the correlation strength is strongly dependent on $f_{\rm com/indep}$, we find that, a smaller $f_{\rm com/indep}$, decreasing from the fiducial value of 1.5 to 1.0, may be possible for NGC 4151 as shown in Figure~\ref{fig:ngc4151_better}.

As shown in Section~\ref{sec:diff_rin}, the model predicted UV to X-ray lag decreases with increasing the inner edge of the cold accretion disk.
In Figure~\ref{fig:large_UV2Xray}, we also show the medians of ideal simulations (the dashed lines) implied by a truncated disk ($r_{\rm in} > r_{\rm ISCO}$).
On the left panels of Figure~\ref{fig:large_UV2Xray} for NGC 4151 with $r_{\rm ISCO} = 1.61~r_{\rm g}$, the selected $r_{\rm in} = r_{\rm ISCO} f^{14}_{\rm rbr} \simeq 6.1~r_{\rm g}$ is equivalent to the ISCO radius of a Schwarzschild BH. Comparing to data, the smaller UV to X-ray lags implied by $r_{\rm in} \simeq 6.1~r_{\rm g}$ may indicate that the cold disk should extend more inwards and so a Kerr BH is more preferred for NGC 4151.
On the right panels of Figure~\ref{fig:large_UV2Xray} for NGC 4593 with $r_{\rm ISCO} = 6~r_{\rm g}$, $r_{\rm in} = r_{\rm ISCO} f^{15}_{\rm rbr} \simeq 25.1~r_{\rm g}$ is selected for a significant change of the UV to X-ray lag, while still smaller than the typical radius of the UV emission regions, $r_{\rm UV}$.
Although the cold disk for NGC 4593 is possibly truncated as well, the single measurement of the UV to X-ray lag and its large uncertainty hinder us rejecting the baseline model with $r_{\rm in} = r_{\rm ISCO}$ for NGC 4593.

On the correlation-wavelength relation predicted by our models in the top panels of Figure~\ref{fig:large_UV2Xray}, there is a prominent dip at $\sim 10-100$~\angstrom. Around these wavelengths, the flux contributions from disk and corona are comparable, and the Wien steep portion of disk emission is supplied by only a few disk zones, thus poorly correlates with variation at other wavelength. Consequently, the scatter of lag is larger. 
It is surprising to note that on the correlation-wavelength relation our current toy model predicts a quite similar shape to that observed around $\sim 1$ keV for NGC 4151 with $r_{\rm in} = r_{\rm ISCO} = 1.61~r_{\rm g}$. However, currently we do not count this as an additional plus of our toy model, since our model simply deals with the corona by assuming a single power-law emission and has not considered the soft excess as well as the reflection component, which need to be improved in the future.

All in all, interestingly, considering the scatter, our model predicts the lag-wavelength relation in agreement with data of NGC 4151 and NGC 4593, which seriously challenges the reprocessing model.

\subsection{A truncated cold disk: NGC 5548}\label{sect:ngc5548}

\begin{figure}[t!]
\centering
\includegraphics[width=0.4\textwidth]{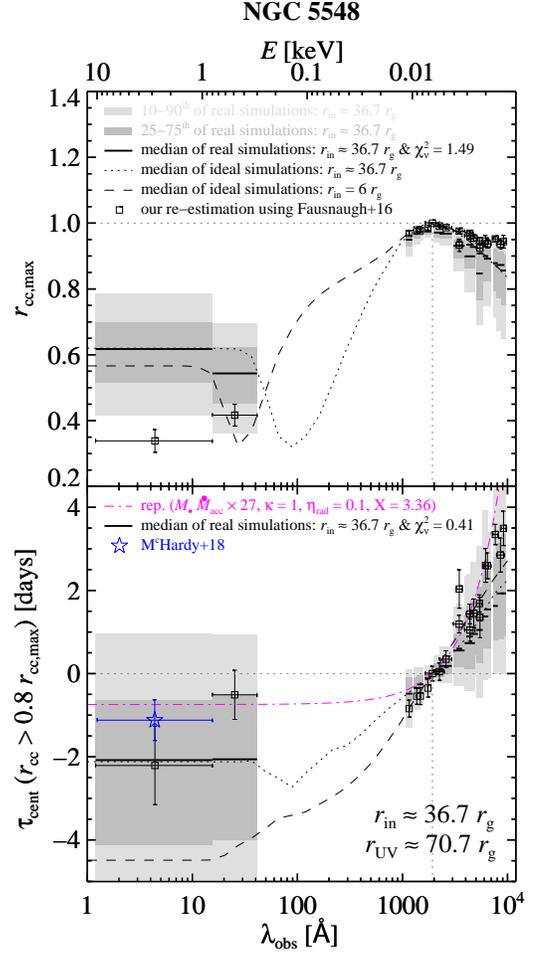}
\caption{In analog to Figure~\ref{fig:large_UV2Xray}, but for NGC 5548. 
Here the baseline model for NGC 5548 has $r_{\rm in} \simeq 36.7~r_{\rm g}$, compared to the one with $r_{\rm in} = r_{\rm ISCO} = 6~r_{\rm g}$ (dash lines).
A potential negative UV to X-ray lags (i.e., UV leads X-ray) for NGC 5548 is predicted by our model as a result of its truncated UV emitting region with $r_{\rm in}$ approaching $r_{\rm UV}$(1928~\angstrom).
}\label{fig:ngc5548}
\end{figure}

\begin{figure}[t!]
\centering
\includegraphics[width=0.4\textwidth]{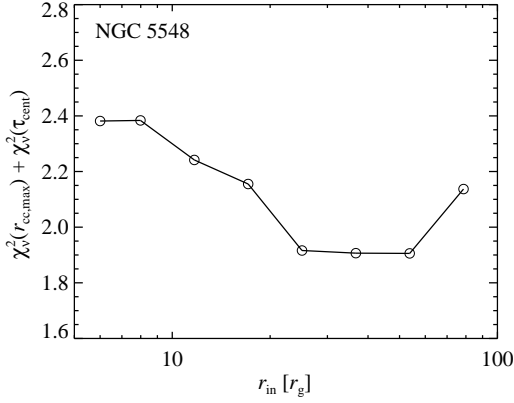}
\caption{ 
Using $\chi^2_\nu(r_{\rm cc,max}) + \chi^2_\nu(\tau_{\rm cent})$ to assess the differences between our model with distinct $r_{\rm in}$ and observations for NGC 5548 indicates its thin disk is potentially truncated at $r_{\rm in} \simeq 25-50~r_{\rm g}$.
}\label{fig:ngc5548_grid_fit}
\end{figure}

\begin{figure}[t!]
\centering
\includegraphics[width=0.4\textwidth]{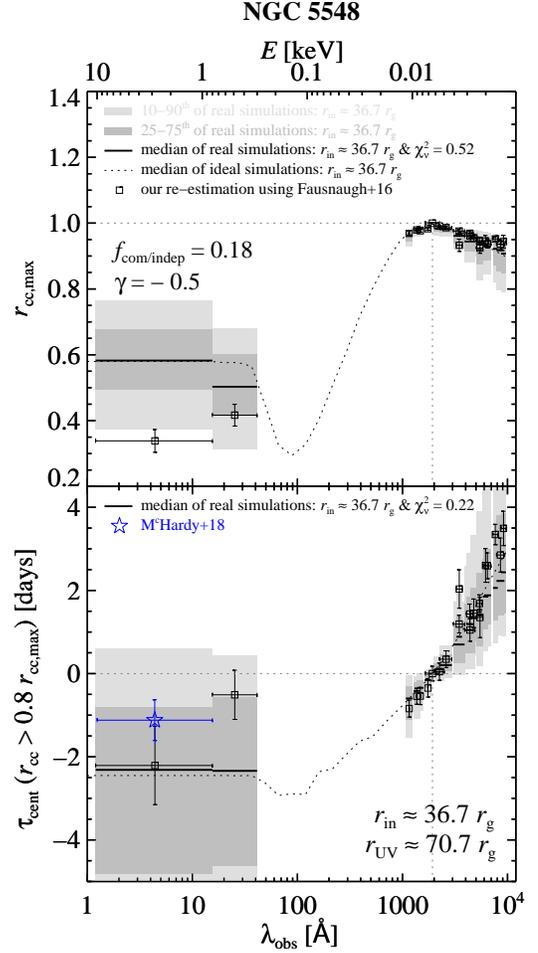}
\caption{ 
In analog to Figure~\ref{fig:ngc5548} for NGC 5548 with the same $\alpha~(=1)$ and $r_{\rm in}$ but different radius-dependence of $\sigma_{\rm s,com}(r) = f_{\rm com/indep} \times \sigma_{\rm s,indep}(r) \times (r/r_{\rm g})^{\gamma+\alpha/2}$. Here the illustrated model has $f_{\rm com/indep} = 0.18$ and $\gamma = -0.5$, compared to the baseline model with $f_{\rm com/indep} = 1.5$ and $\gamma = -1.0$ presented in Figure~\ref{fig:ngc5548}.
}\label{fig:ngc5548_better}
\end{figure}

\begin{figure}[t!]
\centering
\includegraphics[width=0.4\textwidth]{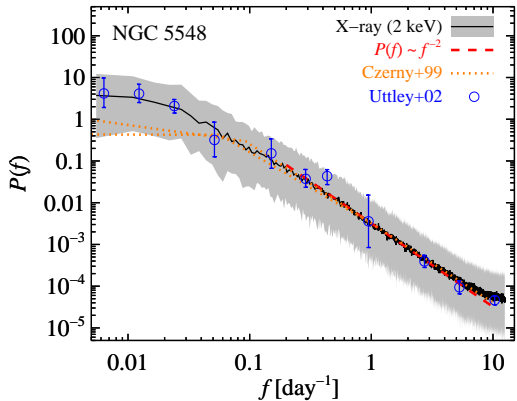}
\caption{ 
The median X-ray power density spectrum (2 keV; black solid line) and its 25-75$^{\rm th}$ percentile range (gray region) inferred from the baseline model for NGC 5548 with $r_{\rm in} \simeq 36.7~r_{\rm g}$, compared to the power density spectrum of $\sim f^{-2}$ at high frequency (red dashed line), those deduced from the measured variances at three timescales by \citet[][orange dotted lines]{Czerny1999}, and that measured by \citet[][open blue circles]{Uttley2002}. Those later power density spectra have been normalized at one day to the median one inferred from our baseline model.
}\label{fig:ngc5548_psd}
\end{figure}

In analog to Figure~\ref{fig:large_UV2Xray}, the correlation- and lag-wavelength relations for NGC 5548 are shown in Figure~\ref{fig:ngc5548}, where the data are from \citet{McHardy2018} and our own measurements using the X-ray/UV/optical light curves of \citet{DeRosa2015}, \citet{Edelson2015}, and \citet{Fausnaugh2016}.

A quite large UV to X-ray lag is predicted by our model if the inner edge of the cold disk can extend downward to the ISCO radius for NGC 5548, as shown by the dashed line for the median lag of 200 ideal simulations in the bottom panel of Figure~\ref{fig:ngc5548}. Instead, the current data of the UV to X-ray lags potentially indicate in our model a truncated cold disk for NGC 5548. To search for the truncation radius, we set up a grid of $r_{\rm in}$ and assess the differences between our model and observations using $\chi^2_\nu(r_{\rm cc,max}) + \chi^2_\nu(\tau_{\rm cent})$. As shown in Figure~\ref{fig:ngc5548_grid_fit}, the thin disk of NGC 5548 is potentially truncated at $r_{\rm in} \simeq 25-50~r_{\rm g}$. 
By virtual of future more solid estimates on the UV to X-ray lags for NGC 5548, it would be possible to precisely determine its $r_{\rm in}$.

Accordingly, the baseline model for NGC 5548 with $r_{\rm in} \simeq 36.7~r_{\rm g}$ is shown in Figure~\ref{fig:ngc5548}, and the baseline model for NGC 5548 is in quite well agreement with observations. 
Nevertheless, the baseline model for NGC 5548 seemingly implies a slightly stronger correlation in the X-ray while a weaker correlation in the UV/optical.
Interestingly, a even better agreement on the inter-band correlations can be expected if the relative contribution of the large-scale common variation over the local independent variation is smaller in the X-ray emitting regions while larger in the optical emitting regions. This scheme is easily achievable, for example, by changing the parameter $\gamma$. As shown in Figure~\ref{fig:ngc5548_better}, the agreement is indeed improved if $\gamma$ increases from -1.0 used in the baseline model to -0.5. Note that $f_{\rm com/indep}$ accordingly decreases from 1.5 to 0.18 in order to get the same $\sigma_{\rm s,com}$ at 70 $r_{\rm g}$, which is the typical UV (1928~\angstrom) emitting radius for NGC 5548.

Comparing to sources shown in Figure~\ref{fig:large_UV2Xray}, the scatter of lag at $\lesssim 100$~\angstrom~for NGC 5548 in Figure~\ref{fig:ngc5548} is more significant, primarily attributed to its large $r_{\rm in}$ comparable to $r_{\rm UV}$ (see also the middle panel of Figure~\ref{fig:diff_r_in}). 
Once the truncated radius approaches to the typical UV emitting regions, the timescales of temperature fluctuations inducing both the UV and X-ray variations are similar and therefore may result in zero or even negative UV to X-ray lags, i.e., the UV leading the X-ray, owing to the random turbulence.
The negative UV to X-ray lag is an intriguing prediction of our model (see Section~\ref{sect:randomness} for further discussion).

As mentioned in Section~\ref{sect:sim_app}, the median X-ray power density spectrum inferred from our model is qualitatively consistent with that found by \citet{Czerny1999}. In Figure~\ref{fig:ngc5548_psd}, we quantitatively compare them using the baseline model for NGC 5548 with $r_{\rm in} \simeq 36.7~r_{\rm g}$. The fact that the agreement at high frequency is excellent is not surprising because we have assumed a power density spectrum of $\sim f^{-2}$ to model the X-ray fluctuations and \citet{Czerny1999} have also assumed almost the same slope to deduce the power density spectrum. 
Instead, at low frequency, although somewhat higher than that deduced by \citet{Czerny1999}, the median power density spectrum implied by our baseline model is well consistent with that directly measured by \citet{Uttley2002}.

\subsection{The intermediate-mass BH candidate: NGC 4395}\label{sect:ngc4395}

\begin{figure*}[t!]
\centering
\includegraphics[width=0.4\textwidth]{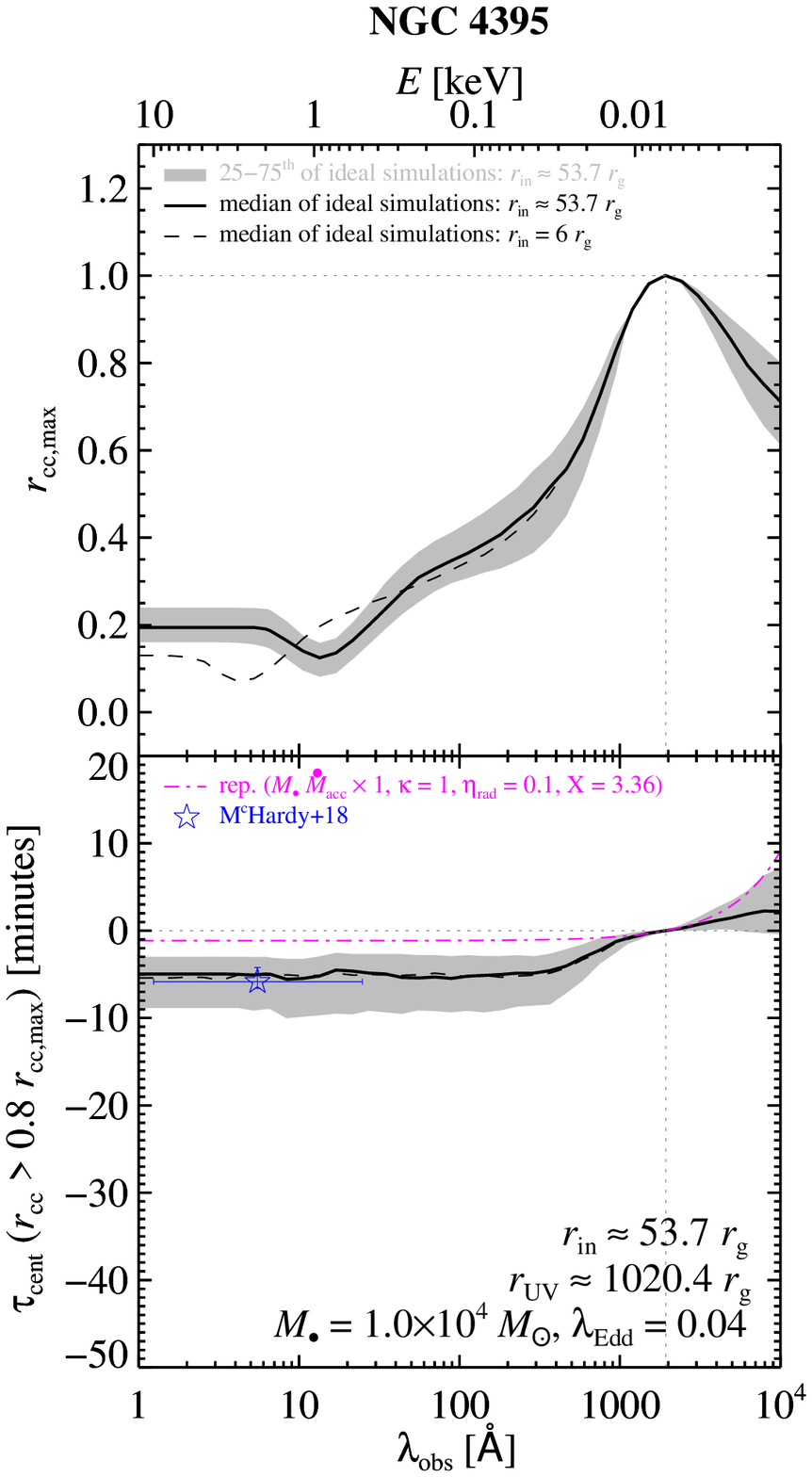}
\includegraphics[width=0.4\textwidth]{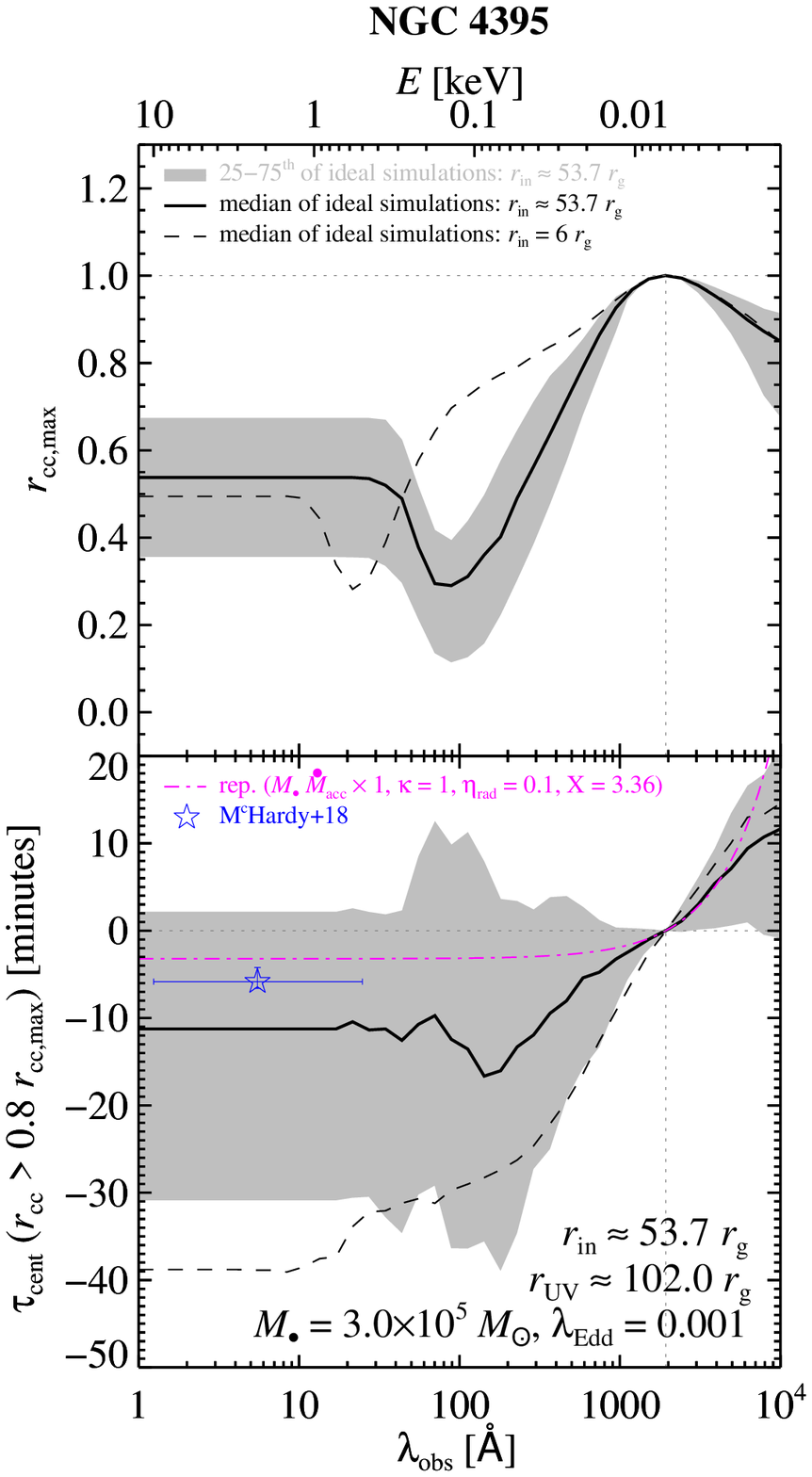}
\caption{In analogy to Figure~\ref{fig:large_UV2Xray}, but for NGC 4395 with a small BH mass (left panels; $M_\bullet = 1.0 \times 10^4~M_\sun$ and $\lambda_{\rm Edd} = 0.04$; \citealt{Woo2019}) and a large BH mass (right panels; $M_\bullet = 3.0 \times 10^5~M_\sun$ and $\lambda_{\rm Edd} = 0.001$; \citealt{Peterson2005}). 
Owing to their unavailable light curves, only the median (black solid line) and 25-75$^{\rm th}$ (gray region) percentile ranges of 200 ideal simulations with $r_{\rm in} \simeq 53.7~r_{\rm g}$ are shown to compare with the data from \citet[][blue open star]{McHardy2018}.
The dashed lines show the median results of 200 ideal simulations with $r_{\rm in} = r_{\rm ISCO} = 6~r_{\rm g}$.
The lag-wavelength relation implied by the reprocessing model (magenta dot-dashed lines) are also shown for comparison.
}\label{fig:ngc4395}
\end{figure*}

Since its discovery \citep{FilippenkoSargent1989}, NGC 4395 is one of the few intermediate-mass BH candidates \citep{FilippenkoHo2003,Dong2012,Greene2019}. Based on the reverberation mapping of the C IV $\lambda$1549 line, \citet{Peterson2005} estimate its BH mass as $M_\bullet \simeq (3.6 \pm 1.1) \times 10^5~M_\sun$ (see also, \citealt{denBrok2015}, \citealt{Brum2019}). However, its mass has recently been updated by \citet{Woo2019} to a remarkably lower value of $\simeq 9.1^{+1.5}_{-1.6} \times 10^3~M_\sun$, based on the narrow-band reverberation mapping of the H$\alpha$ line.

Thanks to its relatively low-mass BH and therefore expected small characteristic timescales of variations, suitable for intensive monitoring campaigns, NGC 4395 has been observed by {\it XMM-Newton} \citep{Vaughan2005,McHardy2016}, {\it Chandra}+{\it HST} plus ground optical telescopes \citep{Desroches2006,ONeill2006}, and {\it Swift} \citep{Cameron2012}. 
Recently, \citet{McHardy2016} state a $UVW1$ to 0.5-10 keV lag of $\sim 7.9^{+0.8}_{-1.6}$ minutes.
Since the light curves of NGC 4395 are generally unavailable, we only show this up-to-date UV to X-ray lag \citep[cf. Table~3 of][for the corrected value relative to the {\it UVW2} band]{McHardy2018} in the bottom panels of our Figure~\ref{fig:ngc4395}.

Considering the ongoing debate on the BH mass estimate, we show in Figure~\ref{fig:ngc4395} the correlation- and lag-wavelength relations predicted by our model with a small BH mass of $M_\bullet = 10^4~M_\sun$ \citep[cf.][]{Woo2019} and a large BH mass of $M_\bullet = 3 \times 10^5~M_\sun$ \citep[cf.][]{Peterson2005}. The Eddington ratio is then adjusted accordingly given the bolometric luminosity from \citet{McHardy2018}.
In both cases, the baseline models (solid thick lines) have a truncated cold disk with $r_{\rm in} \simeq 53.7~r_{\rm g}$, which is selected to reduce the UV to X-ray lags when the large BH mass is assumed (see the dashed line in the bottom-right panel of Figure~\ref{fig:ngc4395} for $r_{\rm in} = r_{\rm ISCO} = 6~r_{\rm g}$).
Instead, in the bottom-left panel of Figure~\ref{fig:ngc4395}, we find that the lag-wavelength relations are more or less the same for $r_{\rm in} \simeq 53.7~r_{\rm g}$ and $r_{\rm in} = r_{\rm ISCO} = 6~r_{\rm g}$ when the small BH mass is assumed.

Assuming a large BH mass for NGC 4395 \citep{Peterson2005}, \citet{McHardy2018} find that only in NGC 4395 the UV to X-ray lag predicted by the reprocessing model is consistent with that observed (see also the comparison between the magenta dot-dashed line and the blue open star in the bottom-right panel of Figure~\ref{fig:ngc4395}). However, if NGC 4395 holds a BH with mass as low as that reported by \citet{Woo2019}, the reprocessing model would be also challenged by the same puzzling large UV to X-ray lag of NGC 4395 (see the bottom-left panel of Figure~\ref{fig:ngc4395}).
Based on our predictions shown in Figure~\ref{fig:ngc4395}, future intensive and sensitive X-ray/UV/optical monitoring on NGC 4395 would help discriminating its BH mass as well as the truncated radius of the cold accretion disk.

\section{Discussions}\label{sect:discussion}

\subsection{Correlation between the UV to X-ray lag and the broadness of Fe K$\alpha$}\label{sec:FeKa}

\begin{deluxetable*}{lcc ccc}[t!]
\tablenum{2}
\tablecaption{The UV to X-ray lags and the Gaussian width of Fe K$\alpha$ line for the concerned Seyfert galaxies\label{tab:lag_diff}}
\tablewidth{0pt}
\tablehead{
\colhead{} & \colhead{} & \multicolumn{3}{c}{The observed-frame UV ({\it Swift}-{\it UVW2}) to X-ray lags} & \colhead{The Gaussian width} \\
\cline{3-5}
\colhead{} & \colhead{X-ray band} & \colhead{Observed} & \colhead{Reprocessing model} & \colhead{Our model} & \colhead{$\sigma_{\rm Fe~K\alpha}$(6.4 keV)} \\
\colhead{Object} & \colhead{[keV - keV]} & \colhead{[days]} & \colhead{[days]} & \colhead{[days]} & \colhead{[keV]} \\
\colhead{(1)} & \colhead{(2)} & \colhead{(3)} & \colhead{(4)} & \colhead{(5)} & \colhead{(6)}
}
\startdata
NGC 4151 & 5-10   & $3.58^{+0.36}_{-0.46}$                & 0.178                 & $3.32^{+0.71}_{-0.71}$                & $0.39 \pm 0.04$ \\
NGC 4593 & 0.5-10 & $0.66 \pm 0.15$                       & 0.089                 & $0.87^{+0.40}_{-0.22}$                & $0.30^{+0.13}_{-0.07}$ \\
NGC 5548 & 0.8-10 & $1.12 \pm 0.49$                       & 0.281                 & $2.08^{+2.04}_{-1.44}$                & $0.035 \pm 0.02$ \\
NGC 4395 & 0.5-10 & $4.05^{+0.54}_{-1.13} \times 10^{-3}$ & $8.65 \times 10^{-4}$ & $3.53^{+2.81}_{-1.43} \times 10^{-3}$ & $0.09 \pm 0.05$ \\
\enddata
\tablecomments{\footnotesize 
Column 1: Name of object. 
Column 2: The energy range of the X-ray band, relative to which the UV to X-ray lag is measured.
Column 3: The observed-frame UV ({\it Swift}-{\it UVW2}) to X-ray (the corresponding band in Column 2) lags reported by \citet[][cf. their Table 3]{McHardy2018}. These UV to X-ray lags for NGC 4151, NGC 4593, NGC 4395, and NGC 5548 are from \citet{Edelson2017}, \citet{McHardy2018}, \citet[corrected to the {\it Swift}-{\it UVW2} band,][]{McHardy2016}, and \citet{Edelson2015}, respectively.
Column 4: The reprocessing model lags reported by \citet{McHardy2018} and rescaled by $(M^2_\bullet \lambda_{\rm Edd})^{1/3}$ given our adopted $M_\bullet$ and $\lambda_{\rm Edd}$ (see Table~\ref{tab:model_par}).
Column 5: The observed-frame UV to X-ray lags implied by our baseline model with parameters tabulated in Table~\ref{tab:model_par}. The median and 25-75$^{\rm th}$ percentile ranges inferred from 200 simulations are presented. Only ideal simulations are applied for NGC 4395, whose light curves are unavailable in literature, while ``real'' simulations are applied for the other sources.
Column 6: The Gaussian widths with 90\% confidence level of the 6.4 keV Fe K$\alpha$ line for NGC 4151, NGC 4593, NGC 4395, and NGC 5548 are from \citet[{\it NuSTAR} + {\it Suzaku},][]{Keck2015}, \citet[{\it XMM-Newton} + {\it NuSTAR},][]{Ursini2016b}, \citet[{\it Suzaku},][]{Iwasawa2010}, and \citet[{\it Suzaku},][]{Brenneman2012}, respectively.
}
\end{deluxetable*}

\begin{figure}[t!]
\centering
\includegraphics[width=0.45\textwidth]{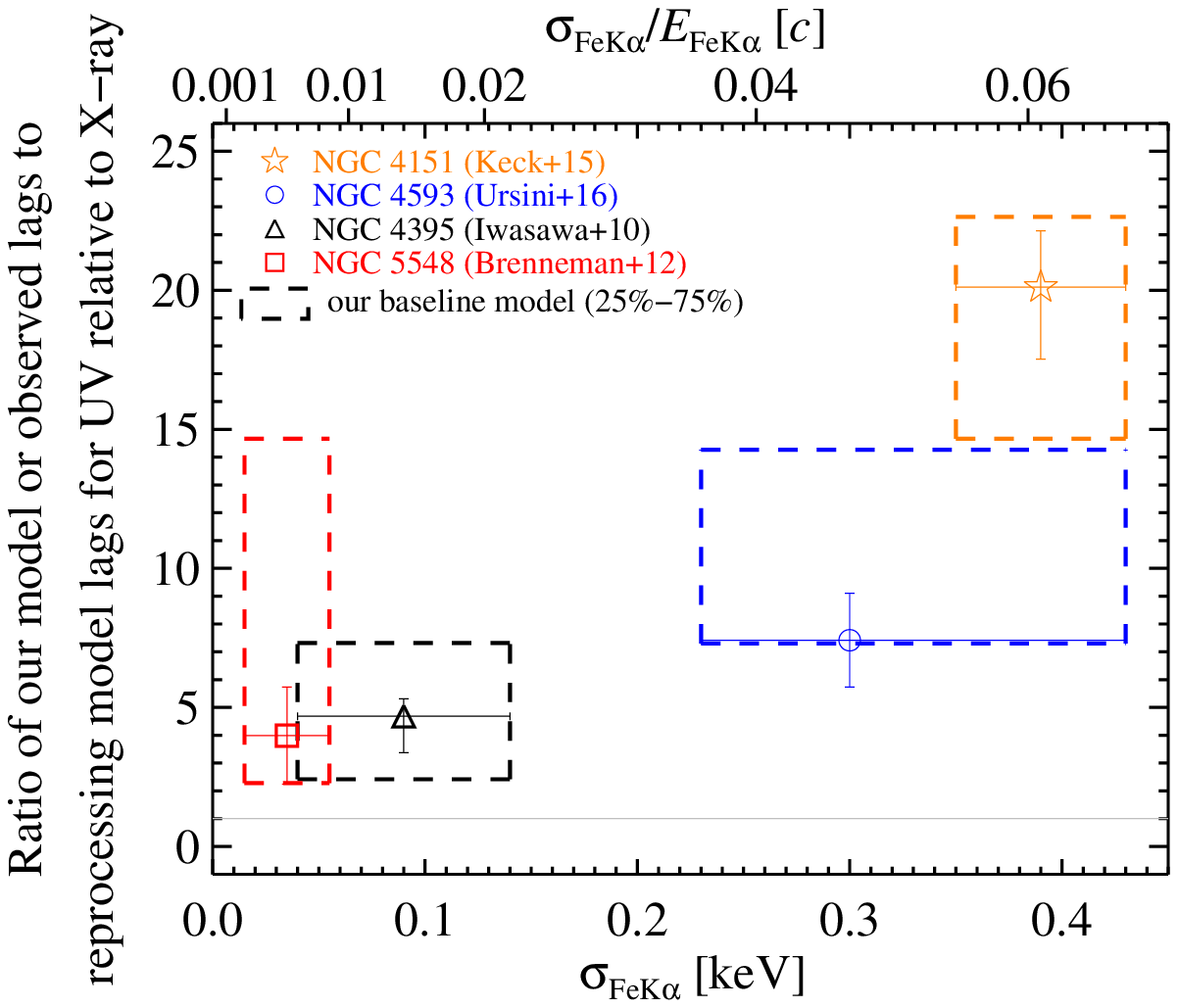}
\caption{
The ratio of our model (dashed-line enclosed regions for the 25-75$^{\rm th}$ percentile ranges) or observed (symbols) to the reprocessing model lags for UV relative to X-ray. The ratio is plotted as a function of the Gaussian width of Fe K$\alpha$ line. (see Section~\ref{sec:FeKa} and Table~\ref{tab:lag_diff})
}\label{fig:xray2uv_lag}
\end{figure}

Following \citet{McHardy2018} and tabulated in Table~\ref{tab:lag_diff}, we plot in Figure~\ref{fig:xray2uv_lag} the ratio of observed to reprocessing model UV to X-ray lags for these Seyfert galaxies.
We directly adopt the observed and reprocessing model lags given by \citet{McHardy2018}, the latter of which is rescaled according to our assumed BH mass and Eddington ratio tabulated in Table~\ref{tab:model_par}.
The larger (than unity) the ratio, the reprocessing model is more challenged. Similar to illustrated in Figures~\ref{fig:large_UV2Xray}, \ref{fig:ngc5548}, and \ref{fig:ngc4395}, our toy model presented in this work could yield UV to X-ray lags well consistent with observations.

As introduced in Section~\ref{sec:diff_rin}, relatively larger UV to X-ray lag is expected for AGN with smaller inner edge of cold disk and thus more compact corona. A smaller inner edge of cold disk would give rise to a broader Fe K$\alpha$ line around $\sim 6.4$ keV.

We note that a relativistic broadened Fe K$\alpha$ line is clearly resolved in NGC 4151 by \citet[][$\sigma_{\rm FeK\alpha} = 0.39 \pm 0.04$ keV]{Keck2015} using {\it NuSTAR}+{\it Suzaku} (see also \citealt{Yaqoob1995,Wang1999,Zoghbi2012,Beuchert2017}; however, see \citealt{Zoghbi2019}). 
A broad component of Fe K$\alpha$ line is statistically required in NGC 4593 by \citet[][$\sigma_{\rm FeK\alpha} = 0.30^{+0.13}_{-0.07}$ keV]{Ursini2016b} using {\it XMM-Newton}+{\it NuSTAR}.

Meanwhile, the lack of a broad Fe K$\alpha$ line is reported for NGC 5548 by \citet[][$\sigma_{\rm FeK\alpha} = 0.035 \pm 0.02$ keV]{Brenneman2012} using {\it Suzaku} and is confirmed by \citet[][$\sigma_{\rm FeK\alpha} \sim 0.04$ keV from their Figure~7]{Cappi2016} using {\it XMM-Newton}+{\it NuSTAR}. Note that, fitting the X-ray spectral data of NGC 5548, \citet{Brenneman2012} estimate a disk truncation radius of $r_{\rm in} \sim 12-90~r_{\rm g}$, which is consistent with our chosen to match the observed UV to X-ray lags.
For NGC 4395, \citet{Iwasawa2010} only detect a weak Fe K$\alpha$ line with $\sigma_{\rm FeK\alpha} = 0.09 \pm 0.05$ keV using {\it Suzaku}. 

Interestingly, Figure~\ref{fig:xray2uv_lag} illustrates a feasible correlation between the ratio and the broadness of Fe K$\alpha$ line, where sources with relatively larger UV to X-ray lags do have more relativistically broadened Fe K$\alpha$ lines, indicative of more compact corona and smaller innermost disk radius.

If confirmed with more X-ray and UV monitoring campaigns, this interesting discovery would provide a new probe to the inner disk/corona, and potentially promote the studies of the broad Fe K$\alpha$ line and the measurements of supermassive BH spin.

\subsection{Randomness of the UV to X-ray lag and inter-band correlations?}\label{sect:randomness}

As illustrated in Figure~\ref{fig:ngc5548}, the large scatter of the UV to X-ray lag for NGC 5548 implies that sometimes there could be no lag between UV and X-ray or even X-ray lagging behind UV, akin to those found in Mrk 817 \citep{Morales2019} and Mrk 509 \citep{Edelson2019}, respectively. This fluctuating property may also account for the failure of recovering lags for the vast majority of AGNs seen in most monitoring seasons by \citet{Yu2018}, as well as for the uncorrelated optical and UV flux variations implied by \citet{Xin2020}.

A tight correlation between the UV/optical continuum variations and the UV/optical broad emission line variations is the foundation of the reverberation mapping technique \citep{BlandfordMcKee1982}, which assumes (1) the UV/optical broad emission lines arise from large-scale clouds photoionized by the largely unobservable intense ionizing continuum or extreme-UV radiation originating from the very inner regions of accretion disk, and (2) the variation of the UV/optical continuum radiation is a good proxy of that of the largely unobservable extreme-UV continuum radiation.
Observationally, this correlation generally holds. However, analyzing the intensive UV/optical monitoring on NGC 5548 with a time baseline of $\sim 180$ days in 2014, \citet{Goad2016} report a transient anomalous phenomenon lasting $\sim 65-70$ days during the campaign. This transient anomalous phenomenon, called the UV anomaly, is indicated by a decorrelation between the UV continuum and emission line variations, characterized by a significant decrease of both flux and variation amplitude of the broad emission lines. Such an anomaly is also found in some emission lines of three high-$z$ luminous quasars by \citet{Lira2018}. Recently, it has been claimed to be very common and probably in every object by \citet{Gaskell2019}.

Current explanations for this anomaly include a falling corona \citep{Sun2018b} or an equatorial obscurer (e.g., a disk wind) with increasing density \citep{Dehghanian2019b,Dehghanian2019a}, obscuring the X-ray and extreme-UV emissions then inducing the spectral variations as observed by \citet{Mathur2017}.
Our scenario potentially provides a different explanation for this anomaly. As shown in Figure~\ref{fig:ngc5548}, sometimes the correlation between extreme-UV and UV emissions could be very weak, owing to the randomness of turbulence and, especially, a disk with truncated radius comparable to the typical radius of UV emitting regions. In other words, the aforementioned second assumption in the reverberation mapping technique is sometimes out of reach. Our scenario directly predicts finding this anomaly with higher frequency in AGNs where there are truncated disks.
Detailed studies on the frequency and characteristics of the anomaly will be presented in a separate paper of this series (Z. Y. Cai et al. 2020, in preparation).

\section{Conclusions}\label{sect:conclusion}

In this work, we have further developed the inhomogeneous turbulence model to account for the puzzling large UV to X-ray lags in local Seyfert galaxies.
In our scenario, the UV/optical as well as the X-ray variations are attributed to disk turbulences, so the UV/optical variations are intrinsic to the thermal disk, rather than the reprocessed X-ray irradiation.
When the effect of large-scale turbulence is considered, the lag for the variation at longer wavelength behind that at shorter one is a result of the differential regression capability of local fluctuations when responding to the large-scale fluctuation.

Without the need of light echoing, our inhomogeneous turbulence model can well reproduce the lags and correlations between the X-ray and UV/optical bands. Moreover, relatively larger UV to X-ray lag is expected for AGN with smaller innermost disk radius and thus more compact corona. Interestingly, for the four local Seyfert galaxies studied in this work, sources with relatively larger UV to X-ray lags do have more relativistically broadened Fe K$\alpha$ lines.

Due to the randomness of the turbulence, our scenario predicts that the measured lag could change from one monitoring campaign to another. Therefore, not only the mean of lags measured in several campaigns but also the scatter contain important information on the central engine of AGNs. Once combined with the measurement of Fe K$\alpha$ line, constraints on the physics of the central engine of AGNs as well as the BH spin would be improved.

\acknowledgments

We are grateful to the anonymous referee for many constructive comments and to R. D. Mahmoud for useful discussion.
We are indebted to Ian M. M$^{\rm c}$Hardy for having sent us their light curves of NGC 4593. 
This work is supported by the National Science Foundation of China (grants Nos. 11873045, 11890693, 11421303, and 11603022), National Basic Research Program of China (973 program, grant No. 2015CB857005), and Specialized Research Fund for the Doctoral Program of Higher Education (20123402110030).
J.X.W. thanks support from the CAS Frontier Science Key Research Program (QYZDJ-SSW-SLH006).

\vspace{5mm}

\software{PyCCF \citep{PyCCF2018}}


\bibliographystyle{yahapj}
\bibliography{ms.bbl}

\end{document}